\documentclass[12pt,preprint]{aastex}

\shorttitle{HD 32297 Circumstellar Disk}
\shortauthors{Schneider, Silverstone \& Hines}

\begin{document}

\title{Discovery of a Nearly Edge-On Disk Around HD 32297}

\author{Glenn Schneider\altaffilmark{1}, Murray D.
Silverstone\altaffilmark{1} and Dean C. Hines\altaffilmark{2} }

\altaffiltext{1}{Steward Observatory, The
University of Arizona, 933 N. Cherry Ave., Tucson, AZ 85721}

\altaffiltext{2}{Space Science Institute, 4750 Walnut Street,
Suite 205 Boulder, CO 80301}

\begin{abstract}
We report the discovery of a nearly edge-on disk
about the A0 star HD~32297 seen in light scattered by the disk grains
revealed in NICMOS PSF-subtracted coronagraphic images.  
The disk extends to a distance of at 
least 400 AU (3\farcs3) along its major axis with a 1.1~$\mu$m flux
density of 4.81 $\pm$ 0.57 mJy beyond a radius of 0\farcs 3 from the
coronagraphically occulted star.  The fraction of 1.1~$\mu$m
starlight scattered by the disk, 0.0033 $\pm$ 0.0004, is comparable to
its fractional excess emission at 25 + 60 $\mu$m of $\sim$ 0.0027 as
measured from IRAS data. The disk appears to be inclined 10\fdg5 $\pm$
2\fdg5 from an edge-on viewing geometry, with its major axis oriented
236\fdg5 $\pm$ 1$\degr$ eastward of north.  The disk exhibits unequal 
brightness in opposing sides and a break in the 
surface brightness profile along NE-side disk major axis. Such asymmetries 
might implicate the existence of one or more (unseen) planetary mass companions.
\end{abstract}



\keywords{circumstellar matter --- infrared: stars --- planetary
systems: proto-planetary disks --- stars : individual
(\objectname[HD~32297]{HD~32297})}


\section{Introduction}

After decades of concerted effort applied to understanding the
formation processes that gave birth to our solar system, until recently, the detailed
morphology of circumstellar material that must eventually form planets
has been virtually impossible to discern.  The advent
of high contrast coronagraphic imaging, as implemented with the Hubble
Space Telescope ({\it HST}) instruments, has dramatically enhanced our
understanding of planetary system formation.  Even so, only a handful
of evolved disks ($\gtrsim$ 10$^{6}$ yr) have been imaged and
spatially resolved in light scattered from their constituent grains
 (e.g., Schneider et al.\ 1999; Weinberger et al.\ 1999; Ardila et al.\ 2004;
Krist et al.\ 2005; Kalas et al.\ 2005).
To expand this sample, we are conducting a NICMOS coronagraphic
imaging survey of 26 (typically $\gtrsim$ 10$^{7}$ yr) main sequence stars
with strong thermal-IR excesses (indicative of circumstellar dust) to
provide a larger ensemble of spatially resolved and photometrically
reliable high resolution images of debris disks, and to probe these posited
epochs of planetary system formation and evolution.  Our images 
shed light on the spatial distribution of the dust in disk
systems, revealing disk structures as close as 0\farcs3 from their central stars.
Here, we report our observations of HD~32297, which have
given rise to the the first new circumstellar  disk image to
emerge from our currently executing debris disk candidate survey.

\section{Observations of HD 32297}

HD 32297 (A0; d = 113 pc $\pm$ 12 pc \citep{peri97}; J = 7.69, H =
7.62, K = 7.59 \citep{2MASS}) was highly ranked on our survey
list  because of its far-IR excess emission above the photospheric level, 
 {\it f}$_{ir}$ = ${\it L}_{\it disk,ir}/{\it L}_{*}$ 
 $\gtrsim$  0.0027,  calculated from 25 \& 60~$\mu$m 
fluxes in the IRAS Faint Source  Catalog. 

We conducted NICMOS observations of HD~32297 on 2005 February 24. Following F165M 
($\lambda$$_{eff}$ = 1.674 $\mu$m, FWHM = 0.1985 $\mu$m) target acquisition (ACQ) 
imaging, deep coronagraphic (CORON) images were obtained in the
 F110W ($\lambda$$_{eff}$ = 1.104 $\mu$m, FWHM = 0.5915 $\mu$m) filter at two field
orientation angles differing by 29\fdg2, yielding a total integration
time of 1344 s (Table 1).  Following the coronagraphic imaging at each
field orientation, referenced by {\it HST}/GO 10177 Visit number
(V\#\#) in this Letter, the spacecraft was slewed 2\farcs 83 and the halo of the
core-saturated unocculted (DIRECT) stellar PSF was imaged.

\section{Calibration/Reduction}

{\it BASIC CALIBRATION.} ACQ images were photometrically and
astrometrically calibrated with procedures developed by the NICMOS IDT
(Schneider et al.\ 2002 [=SCH02], \S 9).  The raw MULTIACCUM (CORON
and DIRECT) images were calibrated with the STSDAS CALNICA task (Bushouse
1997; Stobie et al.\ 1998).  ``Synthetic'' dark frames, high S/N
calibration reference flats, and linearity files appropriate for {\it
HST} Cycle 13 were supplied by STScI. 
We augmented the STScI reference flat with data from contemporaneously acquired ACQ 
mode lamp flats  in calibrating the CORON images (SCH02, \S 8) because the imprint of the
coronagraphic hole was shifted w.r.t. our images (Noll et al. 2004, \S 5).

{\it REDUCTION.} The three coronagraphic count rate images in each
visit were median combined, after verifying the stability of both the
target pointing (SCH02, \S1) and of the PSF (Schneider et al\ 2001 [=SCH01],
\S4) by examination  of image pair differences.  The two DIRECT images in 
each visit were similarly processed and then averaged.  The combined
 calibrated images were post-processed to remove well understood image/readout artifacts
and those that arise in the presence of deeply exposed targets, e.g., 
 (SCH02, \S8; SCH01, \S3; Schneider et al.\ 2003, \S 3.1).  The images were
distortion corrected by mapping input pixels with X:Y scales of
(75.950, 75.430) mas/pixel onto a rectilinear grid of 75.430 mas
square pixels using the NICMOS IDT's IDP3 software \citep{idp3}.
Distortion-corrected instrumental count rates were converted to
physical flux densities based upon absolute photometric calibrations
established from the {\it HST} SMOV3B program
(F110W: 1.26 $\mu$Jy/ADU/s, 0$_{mag}$(Vega) = 1775 Jy; 
F165M: 3.00 $\mu$Jy/ADU/s, 0$_{mag}$(Vega) = 1022 Jy;
Rieke 2002).

\section{Image Registration and PSF subtraction}

In each visit, the star's location in the CORON images was determined
using spacecraft offset slew vectors downlinked in the engineering
telemetry applied to star's position measured from the pre-slew ACQ
image.  Stellar image centroids were determined by Gaussian profile fitting, also
using IDP3.  The calibrated CORON images were coaligned by shifting
the V44 image to the location of the V43 image with IDP3's bicubic
sinc-function apodized interpolative resampling.  With the images
registered, simple image subtraction clearly showed the
positive/rotated-negative signature of a nearly edge-on circumstellar
disk.

Coronagraphic reference PSFs were aligned with the HD~32297 images in a
similar manner.  Our reference PSF images are generally drawn from
the null detections in our survey which are all observed in very similarly constructed two-orientation orbits using 
nearly identical exposure sequences.  These
images were reduced, calibrated and processed in the same manner as
the HD~32297 images.  To qualify as a reference PSF, the star must be
at least as bright as the disk target star, and of similar spectral
type to minimize color effects under the F110W filter.  
Additionally, we applied a coronagraphic PSF from HD~9627,
a very bright calibration target of nearly identical spectral type
previously observed coronagraphically and found not to possess
any detectable disk-scattered light.  On 2004 December 22 the {\it HST}
secondary mirror was moved\footnote{http://www.stsci.edu/hst/observatory/focus/mirrormoves.html}
causing small, but noticable, differences in the coronagraphic PSF
structure in subsequent images.  Hence, reference PSF stars for HD
32297 were restricted to stars observed after that date.  For our
HD~32297 PSF subtractions four of our stars met the above criteria: 
HD~9627 (A1V, J= 5.49; Visit 84), HD~142666 (A8V, J=7.35; Visits 2B/2C), 
HD~36112(A3, J=7.22; Visits 31/32), and HD~83870 (F8, J=6.73; Visits 61/62); 
(J magnitudes from Cutri et al.\ 2003).

We separately registered and subtracted all seven PSF star images
after scaling their intensities, separately, from both the the V43 and
V44 images of HD~32297 (see SCH01 for details of process
and error estimation).  We used our DIRECT images of the stars to
estimate the flux density scaling for the reference PSFs.  After
co-aligning the DIRECT images we performed iterative subtractions to
null the star images (masking saturated pixels) and establish the
intensity ratios.  After subtracting each PSF from the V43 and V44 target
observations, 
the ensemble of  PSF-subtracted disk images exhibited some degree of variation.  
The subtractions using HD~142666 and HD~36112, which are anomalously 
red for A stars, produced zonal under- and
over-subtractions at different radii, indicative of target:reference PSF
differences from mis-matched  SEDs.

\section{Disk Images}

Serendipitously, the near-edge-on disk is favorably oriented 
w.r.t.\ the {\it HST} diffraction spikes and is revealed with very similar
morphology in all fourteen PSF subtractions. Importantly, all PSF subtracted 
images from both HD~32297 visits show the same disk structures at both 
field orientations.  The fine structures in the V61 HD~83870 PSF halo and
diffraction spikes were most closely matched to those in the V43 image
of HD~32297. This subtraction, shown in Figure 1, produced the most
artifact free image after alignment ([+0.304, +0.143] pixel
shift) and intensity scaling by 0.440 $\pm$ 0.005. The measures we 
discuss in \S 6 were made from this image.  We assess the systematic
errors in these measures using an ensemble of six PSF subtracted images
from  (V43,V44) and (V61,V62 and V84), all measured identically.  
As a test, we subtracted the selected three reference PSFs from each other, 
and no evidence of any
disk-like ``features'' was found.  We disqualified the HD 142666 and
HD 36112 PSFs subtractions from quantitative error estimation for the
reasons noted in \S 4.

\section{Results}

{\it DISK GEOMETRY.} The HD 32297 disk extends $\gtrsim$ 3\farcs3 (400 AU) 
to the NE of the star, and $\gtrsim$ 2\farcs5  toward the SW (3 $\sigma$ lower limits, 
see Figure 2; the depth of our
integrations limits our sensitivity to low surface brightness flux in the outer 
regions of the disk). Assuming intrinsic circular symmetry, by isophotal ellipse fitting
we find the disk inclined $10\fdg5 \pm 2\fdg5$ from an
edge-on viewing geometry with a major axis  PA = $236\fdg5 \pm 1\degr.$

{\it DISK BRIGHTNESS.} We measured the total (area integrated) disk
brightness, excluding the {\it r} $<$ $0\farcs3$ coronagraphically
obscured region, as 4.81 $\pm$ 0.57 mJy.  We used 
a 7\farcs049 (93 pixel) x 0\farcs682 (9 pixel) rectangular photometric 
aperture centered on the star with its long axis parallel to the 
disk major axis.  The aperture was sufficiently long and wide to
capture all the measurable disk flux, as was tested by incrementally
increasing the aperture  size.  Bifrucating the aperture,  we find no  statistically significant
difference in the total disk flux above (56\%  $\pm$ 12\%)
and below (44\%  $\pm$ 12\%) the disk midplane.

{\it SCATTERING FRACTION.} Because our DIRECT images are
core-saturated, we used HD 32297's spectral type and 2MASS catalog
magnitudes to establish the brightness of the star in the F110W
passband.  
We used the STSDAS CALCPHOT task to
transform 2MASS magnitudes to the NICMOS filter system and found
F110W = 7.71 $\pm$ 0.03 mags, with a corresponding 1.1~$\mu$m stellar
flux density of 1.46 $\pm$ 0.04 Jy.  We tested the robustness of this
procedure by transforming the 2MASS photometry to the NICMOS F165M
band.  The F165M magnitude of HD 32297 measured from our ACQ images is
7.61 $\pm$ 0.03, in agreement with the CALCPHOT prediction of 7.61
$\pm$ 0.05.  We then find the fraction of 1.1~$\mu$m starlight
scattered by the disk,
 {\it f}$_{nir}$ = ${\it L}_{\it disk,nir}/{\it L}_{*,nir}$ 
, for {\it r} $>$ 0\farcs3, is 0.0033 $\pm$
0.0004.

{\it RADIAL SURFACE BRIGHTNESS PROFILE.} We measured the surface
brightness (SB) of the disk along its major axis in both directions
from the star
in square apertures one resolution element (very close to 1.5
pixels) in extent and spaced one resolution element apart to provide
independent measures in every other sample.  The measured flux
densities were converted to SB units, and the major axis radial
profiles of both ``halves'' of the disk are shown in Figure 2.  Our
photometric measurement uncertainties, on spatial scales of a
resolution element, are dominated by residuals from imperfect PSF
subtractions.  Because of the disk's $\sim 10\degr$ inclination, there
is no significant disk flux along, and near, the minor axis beyond the
radius of the coronagraphic hole.  Hence, we measure 1~$\sigma$
background variations along three radials (six points at each radius)
roughly orthogonal to the disk centered on the minor axis (but
avoiding the diffraction spike) to estimate the measurement
uncertainties arising from subtraction residuals at equal radial
distances along the major axis.  These measures are made identically
to the disk flux measures along the major axis.  Beyond 
$\sim$ 2\farcs2, where the read noise becomes significant,
the uncertainties (Figure 2 error bars) grow larger as a fraction of
the disk flux.

{\it DISK ASYMMETRIES.} 
The SW side of the disk (at {\it r} $>$ 0\farcs3) is brighter (3.14 mJy $\pm$ 0.57 mJy)
than the NE  side (1.67 mJy $\pm$ 0.57 mJy). 
In all HD 32297 PSF-subtracted images, at both
field orientations,  the SW side of the disk is brighter near the star
(e.g., at {\it r} $<$ 0\farcs6 in the  V43-V61 major axis radial profile; Figure 2).  
The innermost point in the radial profiles (at 0\farcs35) on
opposite sides of the disk is questionable due to its close proximity
to the edge of the coronagraphic hole, and should be viewed with
caution.  This brightness asymmetry, however, extends several hundred
mas further out and is also seen above and below the mid-plane of the
disk (e.g., Fig 1D). 

The SW and NE profiles are symmetric from 0\farcs5 $<$ {\it r} $<$ 1\farcs7,
while they differ significantly at smaller and larger radii.
The SB profile (SB in mJy arcsec$^{-2}$; {\it r}
in arcsec) of the SW side of the disk at {\it r} $>$ 0\farcs5 is well
represented by a power law: SB(SW)~=~0.455~x {\it r}$^{-3.57}$ with a
goodness of fit R$^{2}$ = 0.996.  Fitting a single power law to the
NE side of the disk: SB(NE)~=~0.45~x~{\it r}$^{-3.2}$ (R$^{2}$ =
0.931) does not do as well.  By inspection, there is an obvious
``break'' in the NE SB profile (Fig 2) at {\it r} = 1\farcs7.  In this
region (1\farcs4 $<$ {\it r} $<$ 2\farcs1) the NE side SB is
systematically lower than the SW side SB. Separately fitting
the  regions on both sides of the break yields:
SB(NE; 0\farcs5 $<$ {\it r} $<$ 1\farcs7) = 0.51 x {\it r}$^{-3.7}$
(R$^{2}$ = 0.940), and SB(NE; 1\farcs7 $<$ {\it r}
$<$ 3\farcs4) = 0.286 x {\it r}$^{-2.74}$ (R$^{2}$ = 0.996).

\section{Discussion}

HD 32297 joins $\beta$~Pic, HR~4796A,  and $\alpha$~Psc~A as an
additional example of a  scattered light disk about an
A-type  star.  One might consider in comparison the
optically thin disk of the Herbig AeBe (B9.5) star HD~141569A, though
this may represent an example of a transitional disk about a star
younger than either $\beta$~Pic or HR 4796A. 
Table 2 summarizes the characteristics of all five disk systems.
 
HD~32297 is located at  the bottom of the A-star main sequence locus in a M$_{V}$ vs. B-V
color magnitude diagram (cf., Jura et al. 1998).  Its  low luminosity 
(M$_{V}$ = +2.88 [+0.27,-0.24]; Perryman et al. 1997) is similar, for its B-V color 
(+0.199  $\pm$ 0.014), as  $\beta$~Pic  and HR~4796A, suggestive of 
youth comparable to these stars with ages of $\sim$ 10$^{7}$ yr \citep{jura}. While HD~32297's 
age is not well constrained, both its  {\it f}$_{nir}$  and {\it f}$_{ir}$ 
are also comparable to these young stars' and to HD~141569A's, but  hundreds of 
times larger than  the much older $\alpha$~Psc~A's.

The HD~32297 disk is intermediate in size between the very 
large $\beta$~Pic disk and the width/radius = 0.2 debris rings circumscribing
HR~4769A and  $\alpha$~Psc~A, but is comparable in size to the 
HD~141569A disk. Dust in the $\beta$~Pic and HR~4796A debris 
systems would likely have dissipated through radiation pressure 
``blow-out'' or infall due to Pointing-Robertson drag (and certainly so 
for $\alpha$~Psc~A), if not replenished (by collisional erosion of planetesimals)
and/or dynamical confinement by resonances with yet undetected
planetary-mass bodies.  The HD~32297 disk cannot be placed
in a proper evolutionary context until its age is better constrained.

The break in HD 32297's radial SB profile could
arise from  a change in the surface density of scattering
grains or differentiation in their properties with distance from the star. 
The latter cannot explain the NE/SW SB profile asymmetries. 
In the cases of HR~4796A and HD~141569A the presence of their M-star 
companions may explain the outer ``truncation'' of their disks (e.g., see Clampin et al.~2003).
The explanations for more complex asymmetries may rest in disk/planet
dynamics.  Evidence for planets in previously imaged debris systems, and
possibly also in HD~141569A, has been offered given the asymmetries in their disks. Azimuthal asymmetries could be explained
by the presence of undetected planets altering an otherwise
azimuthally symmetric dust density distribution by gravitational
perturbations (e.g., Ozernoy et al.~2000). Such a mechanism might
also be responsible for the SB asymmetries in HD~32297's disk.
While still represented by a very small sample,
the occurrence (and diversity) of azimuthal asymmetries in the circumstellar disks
of A-stars seems to be the rule rather than the exception.

\section{Summary}

We have imaged a circumstellar disk about HD 32297 in 1.1~$\mu$m light.  The
disk, inclined 10\fdg5 $\pm$ 2\fdg5 from edge-on with a major axis PA
= 236\fdg5 $\pm$ 1$\degr$, extends at least 3\farcs3 (400 AU) from the
star.  We estimate the 1.1~$\mu$m disk flux density beyond the
0\farcs3 radius region obscured by the coronagraph to be 4.81 $\pm$
0.57 mJy, leading to a 1.1~$\mu$m disk scattering fraction of 0.0033
$\pm$ 0.0004. We find evidence for hemispheric radial
brightness asymmetries in the disk that might be attributed to an
azimuthally anisotropic distribution of the disk grains, possibly due
to the influence of planetary dynamics.

\acknowledgments
We thank our GO 10177 collaborators for their contributions.  Support
for this work was provided by NASA through GO grant 10177 from STScI,
operated by AURA Inc., under NASA contract NAS5-26555.



\clearpage

\begin{deluxetable}{lccllccl}
\tabletypesize{\scriptsize}
\tablecaption{NICMOS OBSERVATIONS OF HD 32297}
\tablewidth{0pt}
\tablehead{
\colhead{Visit} & \colhead{Orient\tablenotemark{a}} & \colhead{Type} & 
\colhead{Filter} & \colhead{Readout Mode\tablenotemark{b}} & 
\colhead{ExpTime\tablenotemark{c}} & \colhead{Exp} & \colhead{Datasets}
}
\startdata
V43 & 201.57 & ACQ  & F165M & ACQ (ACCUM) & 0.305s  &  2 & N8ZU43010 \\
   &        &CORON & F110W & STEP32/NSAMP14 & 224s & 3 & \\
   &        &DIRECT& F110W & SCAMRR/NSAMP21 & 4.06s& 2 & \\
V44 & 230.77 &ACQ   & F165M & ACQ (ACCUM)  & 0.305s & 2 & N8ZU44020 \\
   &        &CORON & F110W & STEP32/NSAMP14 & 224s & 3 & \\
   &        &DIRECT& F110W & SCAMRR/NSAMP21 & 4.06s& 2 & \\
\enddata
\tablenotetext{a}{Position angle of image Y axis (East of North).}
\tablenotetext{b}{See Noll et al. (2004).}
\tablenotetext{c}{Expoure time for each exposure (before combining).}
\end{deluxetable}

\clearpage

\begin{deluxetable}{lccccc}
\tabletypesize{\scriptsize}
\tablecaption{Dusty Disks Around A Stars\label{tbl-2}}
\tablewidth{0pt}
\tablehead{ 
\colhead{} & \colhead{HD 32297} & \colhead{$\beta$~Pictoris$^{(1)}$} & 
\colhead{HR 4796A$^{(2)}$} & \colhead{HD141569A$^{(3)}$}  & \colhead{Fomalhaut$^{(4)}$}\\
}
\startdata
Spectral Type &A0&A5&A0&HAeBe (B9.5)&A3\\
Est. Age (Myr) & ?? &12--20&8&5&200\\
Disk Radius (AU)&400&$\sim$1600$^{(5)}$&70&400&141\\
{\it f}$_{nir}$&0.0033&$\sim$0.003$^{(6)}$&0.0024&0.0025&10$^{-6}$\\
{\it f}$_{ir}$&0.0027&0.0015$^{(7)}$&0.005&0.0084&5x10$^{-5~(7)}$\\
Anisotropies&see \S 6&(1) &(2) &(8) &(4)\\
\enddata
 \tablerefs{(1) Kalas \& Jewitt 1995, (2) Schneider et al 1999, 
 (3) Weinberger et al 1999. (4) Kalas et al 2005. (5) Larwood \& Kalas 2001. (6) estimated from Kalas et al 2000. (7) Decin et al. 2003. (8) Clampin et al. 2003}
\end{deluxetable}

\clearpage
\begin{figure}
\epsscale{1.0}
\plotone{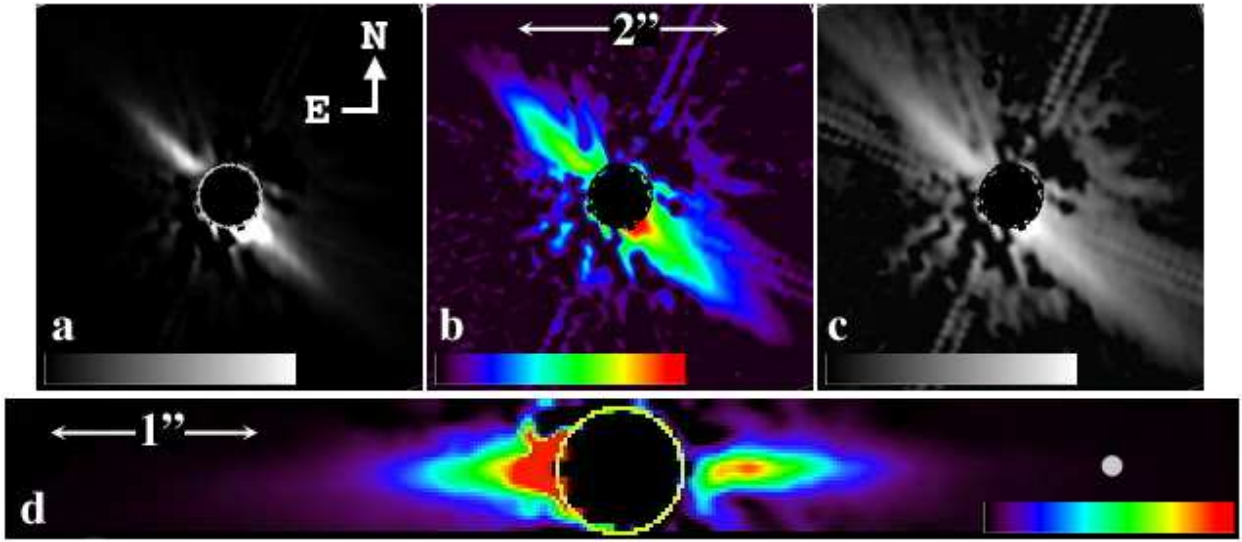}

\caption{V43-V61 PSF-subtracted image of the HD 32297 circumstellar disk.  
Hollow circles indicate  the r = 0\farcs3
coronagraphic hole.  a) Linear display: 0 - 11 mJy arcsec$^{-2}$.  b)
Log$_{10}$ display: [-0.7] to [+1.7] mJy arcsec$^{-2}$.  c) Log$_{10}$ display:
[-1.5] to [+1.5] mJy arcsec$^{-2}$.  d) Horizontally oriented major axis. 
Small gray circle indicates the size of a 0\farcs11
resolution element.  Linear display: 0 to 12 mJy arcsec$^{-2}$.
}
\end{figure}

\clearpage
\begin{figure}
\epsscale{1.00}
\plotone{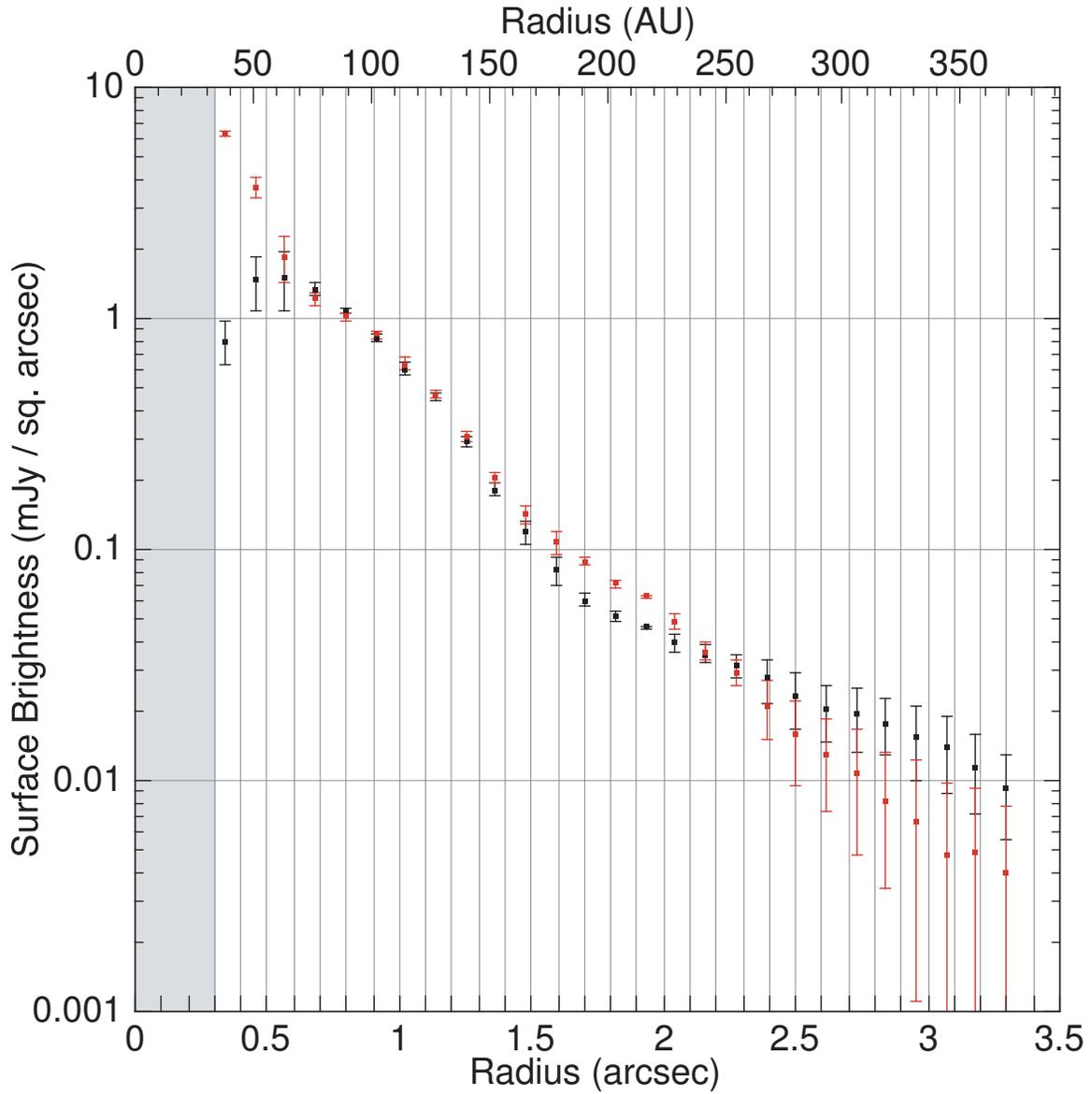} 

\caption{Radial SB profiles along NE (black) and SW
(red) ``halves'' of the HD 32297 disk major axis.  1~$\sigma$ error bars estimate
uncertainties from background variations and do not include the
 $\sim$ 12\% uncertainty in the absolute calibration of the disk flux.
The first resolution element beyond the obscured (gray) region may be affected
by  coronagraphic hole edge effects.
}
\end{figure}

\end{document}